\documentclass[a4paper,11pt]{article}
\pdfoutput=1 

\usepackage{jheppub} 

\usepackage[T1]{fontenc} 

\usepackage{graphicx}
\usepackage{amssymb}
\usepackage{color}
\usepackage{enumerate}
\usepackage{amsmath}
\usepackage{graphics}

\newcommand{\be}{\begin{eqnarray}}
\newcommand{\ee}{\end{eqnarray}}

\newcommand{\bea}{\begin{eqnarray}}
\newcommand{\eea}{\end{eqnarray}}

\begin{document}

	\title{Non-unitarity of Minkowskian non-local quantum field theories}

	\author[a,b,1]{Fabio Briscese,\note{Corresponding author.}}
	\author[c]{Leonardo Modesto}
	
	
	\affiliation[a]{Academy for Advanced Interdisciplinary Studies, Southern University of Science
		and Technology, Shenzhen 518055, China}
	\affiliation[b]{INdAM,  Citt\`{a} Universitaria, P.le A. Moro 5, 00185 Rome, Italy}
	\affiliation[c]{Department of Physics, Southern University of Science and Technology, Shenzhen 518055, China}
	
	\emailAdd{briscesef@sustech.edu.cn, briscese.phys@gmail.com}
	\emailAdd{lmodesto@sustech.edu.cn}

	\abstract{We show that Minkowskian non-local quantum field theories are not unitary. We consider a simple one loop diagram for a scalar non-local field and show that the imaginary part of the corresponding complex amplitude is not given by Cutkosky rules, indeed this diagram violates the unitarity condition. We compare this result with the case of an Euclidean non-local scalar field, that has been shown to satisfy the Cutkosky rules, and we clearly identify the reason of the breaking of unitarity of the Minkowskian theory.  }


	\arxivnumber{XXX}

	\maketitle

\section{Introduction}

The study of  non-local quantum field theory has began a long time ago in the contexts of the Standard Model of particles \cite{efimov,efimov1,efimov10,efimov2,efimov3,efimov4,efimov5,efimov6,efimov7,efimov8,efimov9,efimovscalar} and  stochastic quantization \cite{Namsrai,de la pena}; and it has been revived recently, as it has been realized that non-locality plays an important role at the interplay of quantum field theory and gravitation.

First attempts to achieve a renormalizable theory of gravitation introducing higher derivatives in the Einstein-Hilbert action by Stelle \cite{Stelle}, Krasnikov  \cite{Krasnikov}, and Kuz'min \cite{Kuzmin} date back more than thirty years; see also \cite{FT1,FT2,FT3,FT4,FT5,FT6}. However, it was suddenly realized that, in spite of the fact that such models are renormalizable, they contain unavoidable ghosts. Indeed, higher-derivtive models were abandoned  until it become clear that the occurrence of ghosts could be avoided introducing derivatives of infinite order in a proper manner (or, equivalently, considering non-local interactions) when the quantum theory is defined in Euclidean signature. This has led to the formulation of the so called Non-Local Quantum Gravity (NLQG) \cite{Modesto,Review,ModestoLeslaw,briscese1,briscese2,Buoninfante,Buoninfante2,nonlocal1,nonlocal2,nonlocal3,nonlocal4,nonlocal5,nonlocal6,nonlocal7,nonlocal8,nonlocal9,transplanck,inflation1,inflation2,inflation3,inflation4,inflation5,unitarity1,unitarity2,unitarity3,Yao-dong,nonlocaldesitter,Stability2,Stability1,dona,causality,boos1,boos2,boos3,FT7,FT8,FT9,FT10,FT11,FT12,FT13,FT14}\footnote{We also mention   Lee-Wick quantum gravity as an alternative attempt to remove ghosts in higher derivative theories \cite{HigherDG,Modesto:2015ozb,Modesto:2016ofr,shapiromodesto,LWqg,anselmi,mannheim}.}.

This theory has nice properties both at classical and quantum level. It can be formulated in such a way that all the classical solutions of general relativity  are also solutions of NLQG \cite{Yao-dong}, and  they are as stabile in NLQG as in   Einstein-Hilbert gravity \cite{Stability1,Stability2,nonlocaldesitter}. 
For instance, in NLQG the Minkowski spacetime is stable under any Strongly Asymptotically Flat (SAF) initial data set satisfying a Global Smallness Assumption (GSA) \cite{Stability1}. Of course, NLQG has more solutions than general relativity. In facts, it has been shown that the model has a satisfactory Starobinsky-like inflation \cite{inflation1,inflation2}. The spectrum of scalar perturbations generated during inflation is the same as in the local $R^2$ inflation \cite{Starobinsky}, while  tensor perturbations are affected by the non-locality \cite{inflation3,inflation4,inflation5}; indeed, NLQG is predictive.

At quantum level, the theory is super-renormalizable or even finite \cite{Modesto,ModestoLeslaw,Review}, while it is tree-level indistinguishable from general relativity. Indeed, all the tree-level scattering amplitudes are the same as in the Einstein's theory \cite{dona}, and this implies that the macroscopic causality based on the Shapiro's time delay is  satisfied \cite{causality}. 
Furthermore, in \cite{transplanck} it has been pointed out that relevant NLQG models are asymptotically free in the ultraviolet regime, above the non-locality energy scale $E_{NL}$. This fact has interesting consequences. First, it is impossible to accelerate particles in particle accelerators at energies above  $E_{\rm NL}$, so that, provided that $E_{NL}\leq E_P$, where $E_P$ is the Planck-energy, trans-Planckian energies are unattainable in laboratory experiments. This is due to the fact that, at energies above $E_{NL}$ all interactions are  suppressed and particles  decouple from any device that could accelerate them. This fact, in turns, implies that it is impossible to detect causality violations that occur in non-local theories at small time-scales $\Delta t \sim \ell \sim E^{-1}_{NL}$. In fact, in order to measure such effects, one should be able to test the space-time at scales below $\ell$, and this requires the use of wave-packets tighter than $\ell$, corresponding to particles with energy above $E_{NL}$. Since such energies are unattainable in experiments, causality violations cannot be detected.
Finally, the ultraviolet asymptotic freedom   solves the  cosmological trans-Planckian problem \cite{TransPlanckianProblem1,TransPlanckianProblem2,TransPlanckianProblem3,TransPlanckianProblem4,TransPlanckianProblem5,TransPlanckianProblem6,TransPlanckianProblem8}.
In facts, NLQG is Lorentz invariant and it does not contain extra particles. Moreover, all the fields are asymptotically free above $E_{NL}$, so that quantum gravity corrections are naturally suppressed during all the stages of inflation.

We stress that non-locality  is ubiquitous in  quantum  gravity. In string theory, non-local vertexes of the form $\exp[\Box \, \ell^2]$ appears in interactions \cite{string nonlocality1,string nonlocality2,string nonlocality3,string nonlocality4,string nonlocality5,string nonlocality6}. Furthermore, emergent non-locality at the Planck scale comes from  non-commutative theories \cite{amelino,amelino1}, loop quantum gravity \cite{loop}, asymptotic safety
\cite{asymptotic safety}, and causal sets \cite{causal sets}. Moreover, the trace anomaly induced by quantum  corrections due to conformal fields induces non-local terms in the effective action \cite{anomaly1,anomaly2}. Finally, arguments based on black holes production in scattering processes show that one should expect that non-locality must be  hidden in any quantum gravity  model \cite{addazi}.

As we mentioned above, it has been shown that, introducing non-local interactions, it is possible to construct unitary non-local quantum fields when the theory is defined  in  Euclidean signature \cite{unitarity1,unitarity2,unitarity3}. For Euclidean signature we mean that scattering amplitudes are calculated integrating in $d^4 k_i$ for momenta $k_i\in \mathbb{I}\times \mathbb{R}^3$, where $\mathbb{I}$ is the imaginary axis of the $k^0$ complex plane, assuming that all external energies $E_i$ are purely imaginary. Then, such complex amplitudes are extended by analytic continuation to real energies $E_i$. We remand the reader to \cite{unitarity1,unitarity2,unitarity3} for details, but in Section \ref{Section one loop Euclid} we will give an explicit example calculation of a simple diagram in Euclidean signature for a non-local scalar field.

In this paper we show that, when the theory is defined in Minkowski signature, unitarity is lost. Of course, in Minkowski  signature complex amplitudes are calculated integrating in $d^4 k_i$ for momenta $k_i\in  \mathbb{R}^4$ and assuming that all the external energies $E_i$ are real. As a paradigmatic example, we consider the simple case of a non-local scalar field theory with a Lagrangian
\be
\mathcal{L}_{\phi} = \frac{1}{2} \partial_\mu \varphi \, \partial^\mu \varphi - \frac{1}{2} m^2 \varphi^2 -  \frac{\lambda}{4!} (  e^{ - \frac{1}{2} H[- \sigma (\Box+m^2)] } \varphi )^4 \, .
\label{phin2}
\ee
where $\exp [H]$ is the non-local form factor, and  $\sigma \in \mathbb{R}$ is a parameter with energy dimensions $[\sigma] = - 2$ that fixes  the non-locality length-scale as $\ell_\Lambda = \sqrt{\sigma}$.  Without loss of generality, we set  $H\left[0\right]=0$, corresponding to a rescaling of the coupling constant $\lambda \rightarrow \exp\left(H\left[0\right]\right)$.

\begin{figure}
	\begin{center}
		\hspace{-1cm}
		\includegraphics[height=4cm]{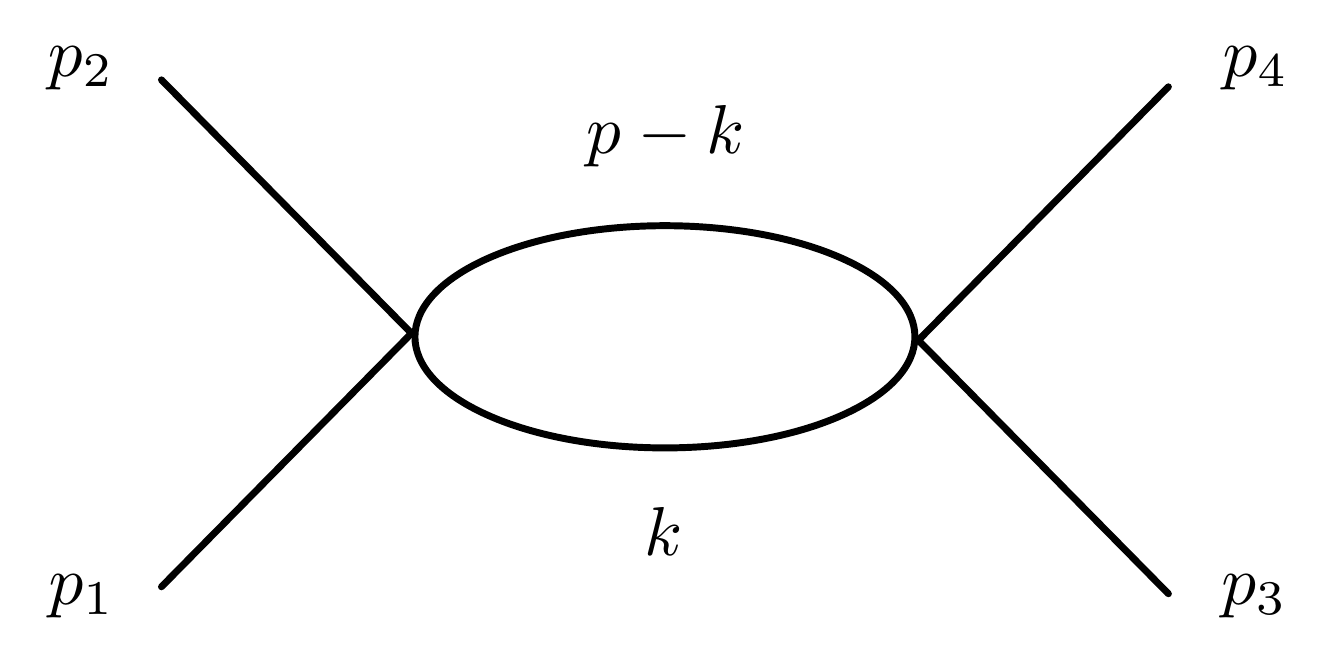}
	\end{center}
	\caption{Four-particle scattering amplitude in $\lambda \phi^4/4!$ theory at one-loop order.}
	\label{Fig2}
\end{figure}

In Section \ref{Section one loop Mink} we will calculate  the scattering amplitude corresponding to the simple one-loop diagram in Fig.\ref{Fig2}, showing that its imaginary part of is not given by the Cutkosky rules. Since such rules are essential for the unitarity, we conclude that the theory in the Minkowskian space is not unitary. For completeness, in Section \ref{Section one loop Euclid} we will calculate the same amplitude when the theory is defined in Euclidean signature, showing that in this case Cutkosky rules give the correct result, indeed the theory is unitary (we remand the reader to \cite{unitarity1,unitarity2,unitarity3} for a complete discussion of the Euclidean case). Finally, we conclude in Section \ref{Section conclusions}.
Before entering  the details of the calculations, in Section \ref{Section unitarity condition} we will review briefly  the unitarity conditions for scattering amplitudes and their relation with Cutkosky rules.

\section{Unitarity condition and Cutkosky rules}\label{Section unitarity condition}

We remind that the unitarity condition $S^\dagger S =1$ for	the scattering matrix $S\equiv 1 + i T$ is usually  expressed in terms of the $T$ matrix as $T - T^\dagger  =  i	\, T^\dagger T$ \cite{peskin,itzykson zuber}. Taking the expectation value of this relation between an
initial incoming state $| a \rangle$ and a  final outgoing state $
\langle b |$ for a given process $a \rightarrow b$, one has

\be T_{ba}  - T^*_{ab}=  i \,  \sum_c T^*_{c b} T_{c a}  \, \ee
Recasting the matrix elements of  $T$  in terms of those
of the invariant scattering amplitude $\mathcal{M}$ as \mbox{
	$T_{ab} = (2 \pi)^4  \, \mathcal{M}_{ab} \,	\delta^{(4)}(\sum_i p_i - \sum_f p_f )$}, where $p_i$ and $p_f$ are the initial an final external momenta, the unitarity
condition can be expressed as

\be\label{unitarity condition M}
&&-i \left(\mathcal{M}_{ba} - \mathcal{M}^*_{ab}\right) = 2 i \,\textit{Im}\left\{\mathcal{M}_{ba}\right\} = \sum_c \,\, \mathcal{M}^*_{c
	b} \,\,\mathcal{M}_{c a} \,\, (2\pi)^4\,\, \delta^{(4)}(p_c-p_a) \, , \label{unitarityM}
\ee 
where $\mathcal{M}_{ba} = \langle b |
\mathcal{M} | a \rangle$ is the sum of all the connected amputated diagrams for the process $a \rightarrow b$ ( see \cite{peskin,itzykson zuber} for a review), and $\textit{Im}\left\{\mathcal{M}_{ba}\right\}$ is its immaginary part. Note that we have neglected a global $\delta^{(4)}(\sum_i p_i - \sum_f p_f )$ multiplying both sides of Eq. (\ref{unitarity condition M}). The sum in $c$ is made on all possible physically admissible intermediate states, i.e., on those states that give nonzero amplitudes $\mathcal{M}_{bc} = \langle b |
\mathcal{M} | c \rangle$ and $\mathcal{M}_{ac} = \langle a |\mathcal{M} | c \rangle$. Of course, such sum becomes an integral when intermediate states $c$ form a continuum set.

The interpretation of (\ref{unitarity condition M}) is that when the energy of the initial state $a$ reaches the threshold of production of intermediate real states $c$, the imaginary part of amplitude $\mathcal{M}_{ba}$   has a  discontinuity corresponding to a branch-cut singularity. In the case of local quantum fields, the singularities of the amplitude are given by the Landau equations \cite{landau},  which are obtained imposing that two or more propagators in the scattering amplitude go on-shell simultaneously; see \cite{itzykson zuber} for a review of the Landau equations. In \cite{unitarity1,unitarity2,unitarity3} it has been shown that in non-local theories the scattering amplitudes have the same singularities of the corresponding local theories. This is due to the fact that the propagators have the same poles as in the local case. Therefore, the non-locality does not change  the singularity structure of the amplitudes.

A crucial step to prove unitarity is to establish the validity of Cutkosky rules \cite{cutkosky}, that state that the l. h. s. of (\ref{unitarity condition M}) is obtained replacing each on-shell propagator in the integral expression of the complex amplitude $\mathcal{M}_{ba}$ with a delta-function, according to the following prescription:

\be\label{cutkosky rule}
\frac{1}{p^2_i -m^2 + i \epsilon}\longrightarrow (-2\pi i) \delta\left(p^2_i - m^2 \right) \, .
\ee
The expression obtained in this way is usually referred as a cut diagram, since it is graphically represented by the same diagram as $\mathcal{M}$ in which the lines corresponding to the on shell propagators are cut  (see \cite{peskin,itzykson zuber,cutkosky,landau} for review). 

Unitarity can be proved showing that Cutkosky rules apply to all normal thresholds, that are those singularities such that, cutting the diagram along on-shell propagators, the diagram is divided in two parts. Moreover, one has to show that only singularities corresponding to normal thresholds contribute to the imaginary part  of $\mathcal{M}$. This is necessary, as some Landau poles  are such that,  cutting the on shell propagators, the diagram is not divided in two, and the imaginary part of the amplitude cannot be recast as in (\ref{unitarity condition M}).  Such Landau poles are usually referred as anomalous thresholds. 

In the case of non-local field theories defined in Euclidean signature, it has been proved  \cite{unitarity1,unitarity2,unitarity3} that Cutkosky rules give the correct prescription for the imaginary part of  complex amplitudes in correspondence of normal thresholds, while  anomalous thresholds do not contribute to (\ref{unitarity condition M}). However, it remained an open issue to state whether the theories defined in the Minkowskian space are unitary as well. Note that, since the theory is non-local, integrals performed in Euclidean and Minkowskian domains are not the same, since they are not simply related by a Wick rotation. This is due to the fact that non-locality introduces a form factor in the complex amplitudes that has an essential singularity at infinity in the $k^0_i$ plane, indeed integrals on the infinite arcs cannot be neglected.

In the next section, analysing a simple one loop diagram, we will show that Cutkosky rules are violated when the theory is defined in Minkowski signature.

\section{One-loop diagram in the Minkowskian non-local theory}\label{Section one loop Mink}

In order to prove the loss of unitarity in the non-local Minkowskian theory, in this section we consider  diagram in Fig.\ref{Fig2} and show that the  Cutkosky rules do not give the correct prescription for the imaginary part of the corresponding complex amplitude. In Minkowski signature, this is given by
\begin{equation}\label{amplitude 1loop 0}
\mathcal{M}(\sigma,p,m,\epsilon) = -  \frac{i \lambda^2}{2} \, \int_{
\mathbb{R}^4} \frac{ \, d^4 k}{(2\pi)^4}
\frac{e^{-H\left[\sigma^2\left(k^2-m^2\right)^2\right]}}{k^2 - m^2 + i \epsilon} \frac{e^{-H\left[\sigma^2\left(\left(k-p\right)^2-m^2\right)^2\right]}}{(k-p)^2 - m^2 + i \epsilon} ,
\end{equation}
where   $p = p_1+p_2=p_3+p_4$, and $p_1, p_2, p_3,p_4$ are the external momenta.

The convergence of the integral (\ref{amplitude 1loop 0}) requires that $H$  must be a function of $\sigma^2 \left(k^2-m^2\right)^2$, corresponding to the replacement $H[-\sigma(\Box+m^2)]\rightarrow H[\sigma^2 (\Box+m^2)^2]$ in (\ref{phin2}). $H[z]$ is such that $z^\gamma \exp\left(-H\left[z\right]\right)\rightarrow 0$ when $z\rightarrow + \infty$, for some  $\gamma > 0$\footnote{A simple power counting shows that, under these hypothesis, the non-local propagator converge to zero faster than $k^{-(4\gamma+2)}$ (in this case $z\sim (\Box+m^2)^2 \sim k^4$), and the superficial degree of divergence $\delta$ for an interaction term $(e^{ - \frac{1}{2} H[- \sigma \Box] } \varphi )^n$ is such that \mbox{$\delta< d +V[-d+n (d-4\gamma-2)/2)]-[(d-4\gamma-2)/2]N$}, for a diagram with $V$ vertices and $N$ external lines in $d$ dimensions. In the case of the Lagrangian (\ref{phin2}) one has $n=d=4$ and $\delta < 4-N-4\gamma I$, where $I= (nV-N)/2$ is the number of internal lines of the diagram. Therefore, the diagram corresponding to $N=4$ external lines is convergent, while the amplitude with $N=2$  is convergent for $\gamma >1/6$.}. In analogy with the Euclidean case \cite{unitarity1,unitarity2,unitarity3}, we assume that $\exp\left(H[z]\right)$ is an entire function  (analytic with no poles except for $|z|=\infty$) without zeros at finite $z$.  Also, thanks to Lorentz invariance,  we can set $\vec{p}=0$ in (\ref{amplitude 1loop 0}) without loss of generality.

Since the form factor $\exp\left(H\left[z\right]\right)$ has no zeros  in the finite complex $z$ plane, the poles of the integrand in (\ref{amplitude 1loop 0}) are only those of the two propagators. Therefore, the singularities of the amplitude (\ref{amplitude 1loop 0}) are given by the same Landau poles as in the local theory, corresponding to the case in which the two propagators are on-shell simultaneously or, equivalently, two of the poles of the propagators merge for some value of the momentum $\vec{k}$. Thus, the amplitude (\ref{amplitude 1loop 0}) has the same singularity structure of the local theory. For the Euclidean theory, this is sufficient to ensure the unitarity, while we will se that this is not the case for the Minkowskian theory.

The poles of the first propagator are
\bea\label{poles 1}
\bar{k}^0_{1,2}
= \pm \sqrt{\vec{k}^2 + m^2 - i \epsilon}\equiv
 \pm \omega(\vec{k},\epsilon)\, . 
\eea
and they do not depend on the
external energy $p^0$. The poles of the second propagator are
\bea\label{poles 2}
\bar{k}^0_{3,4} = p^0 \pm
\omega(\vec{k},\epsilon). 
\eea
It is easy to see that, in the limit $\epsilon \rightarrow 0$, the poles $\bar{k}^0_{1}$ and $\bar{k}^0_{4}$ merge  for $|\vec{k}|=\sqrt{(p^0/2)^2-m^2}$, provided tat $p^0 \geq 2m$, pinching the integration contour in te $k^0$ variable, that is the real axis of the complex $k^0$ plane. In analogy with the case of a local scalar theory, this  implies that the amplitude has a branch-cut singularity at $p^0 = 2m$.

\begin{figure}
	\begin{center}
		\hspace{-1cm}
		\includegraphics[height=6cm]{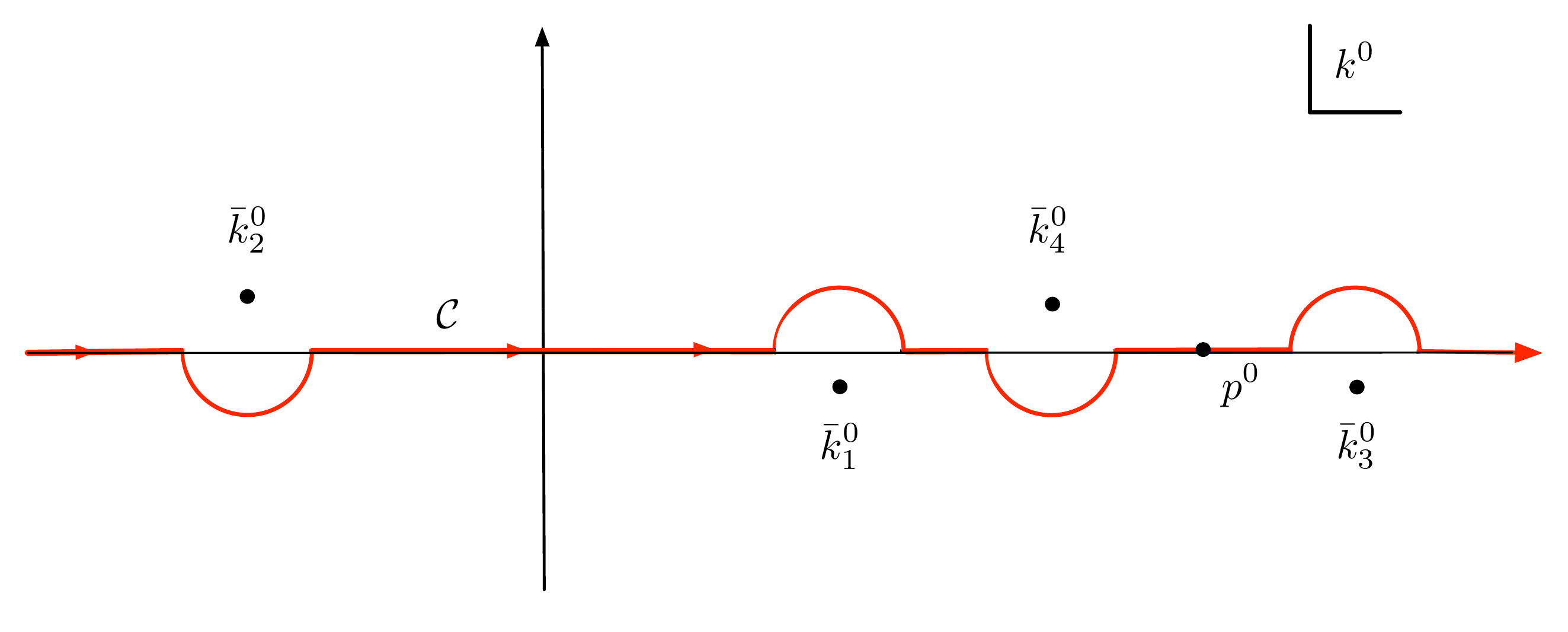}
	\end{center}
	\caption{We plot the poles $\bar{k}^0_1, \bar{k}^0_2,\bar{k}^0_3,\bar{k}^0_4$ in the complex $k^0$ plane. When $\epsilon\rightarrow 0$, such poles become real, and the integration contour is obtained deforming the real axis around  them. The resulting integration contour $\mathcal{C}$ is depicted in red, and the integral in the $k^0$ variable is given by the principal part    plus the contributions of  the  arcs. 
		}
	\label{Fig3}
\end{figure}

When $\epsilon \rightarrow 0$, the poles $\bar{k}^0_1, \bar{k}^0_2,\bar{k}^0_3,\bar{k}^0_4$ become real, and the integral in the $k^0$ variable in (\ref{amplitude 1loop 0}) is performed on the path $\mathcal{C}$ obtained  deforming the real axis around such poles, 
as depicted in in fig. \ref{Fig3}. Therefore, the integral in $k^0$ is  splitted into two contributions: one given by the principal part, the other given by the  infinitesimal arcs. 
Each  arc gives a term $\pm \pi i$ times the residue at the corresponding pole, so that

\bea\label{amplitude 1loop 3}
&& \mathcal{M}(\sigma,p,m) \equiv \lim_{\epsilon \rightarrow 0} \mathcal{M}(\sigma,p,m,\epsilon)  =  \frac{ \lambda^2}{2}    \left[- i \mathcal{P}\left\{\int_{ \mathbb{R}^4} \frac{ \, d^4 k}{(2\pi)^4}
\frac{e^{-H\left[\sigma^2\left(k^2-m^2\right)^2\right]}}{k^2 - m^2} \frac{e^{-H\left[\sigma^2\left(\left(k-p\right)^2-m^2\right)^2\right]}}{(k-p)^2 - m^2 }\right\}
\right.\nonumber 
\\ \nonumber
\\ 
&&+\pi \lim_{\epsilon \rightarrow 0} \int_{ \mathbb{R}^3} \frac{ \, d^3 k}{(2\pi)^4}  \left( Res[\bar{k}^0_2]+Res[\bar{k}^0_4]-Res[\bar{k}^0_1]-Res[\bar{k}^0_3]     \right) \Bigg] ,
\ee
where

\be\label{residues}
&& Res[\bar{k}^0_1]= \frac{1}{2\omega}\frac{e^{-H\left[\sigma^2\left(\left(k-p\right)^2-m^2\right)^2\right]}}{\left(\left(k-p\right)^2-m^2+i \epsilon\right)^2}{\bigg|}_{k^0= \bar{k}^0_1}, Res[\bar{k}^0_2]= -\frac{1}{2\omega}\frac{e^{-H\left[\sigma^2\left(\left(k-p\right)^2-m^2\right)^2\right]}}{\left(\left(k-p\right)^2-m^2+i \epsilon\right)^2}{\bigg|}_{k^0= \bar{k}^0_2}\nonumber\\
&& Res[\bar{k}^0_3]= \frac{1}{2\omega}\frac{e^{-H\left[\sigma^2\left(k^2-m^2\right)^2\right]}}{\left(k^2-m^2+i \epsilon\right)^2}{\bigg|}_{k^0= \bar{k}^0_3}, Res[\bar{k}^0_4]= -\frac{1}{2\omega}\frac{e^{-H\left[\sigma^2\left(k^2-m^2\right)^2\right]}}{\left(k^2-m^2+i \epsilon\right)^2}{\bigg|}_{k^0= \bar{k}^0_4  } \, .
\ee
Note that for $k^0 = \bar{k}^0_2= - \omega$ one has $\left(k-p\right)^2-m^2 \neq 0$ for any $\vec{k}\in \mathbb{R}^3$. In facts, the pole $\bar{k}^0_2$ is always far from $\bar{k}^0_3$ and $\bar{k}^0_4$. That implies that $Res[\bar{k}^0_2]$ has no poles in the integration volume for $\epsilon\rightarrow 0$. In the same way, for $k^0 = \bar{k}^0_3= p^0+ \omega$ one has $k^2-m^2 \neq 0$, and $Res[\bar{k}^0_3]$ has no poles for $\vec{k}\in \mathbb{R}^3$ for $\epsilon\rightarrow 0$.

On the contrary, since $\bar{k}^0_1$ and $\bar{k}^0_4$ coincide in the limit $i\epsilon\rightarrow 0$ for \mbox{ $|\vec{k}|= \sqrt{\left(p^0/2\right)-m^2}\equiv k_p$} when $p^0 \geq 2 m$, $Res[\bar{k}^0_1]$ and $Res[\bar{k}^0_4]$ are singular on the 3-sphere $|\vec{k}|=k_p$. For such terms we use the formula

\begin{equation}\label{delta 1}
\lim_{\epsilon \rightarrow 0} \frac{ 1}{\left(k-p\right)^2 - m^2 + i
	\epsilon}  = \mathcal{P} \left\{ \frac{ 1}{\left(k-p\right)^2 - m^2 }\right\} -\pi i \,
\delta(\left(k-p\right)^2 - m^2) \, ,
\end{equation}
where $\mathcal{P} f(x)$ is the principal part, and this  means that $\int \mathcal{P}f(x) \, dx \equiv \mathcal{P}\int f(x) \, dx$.  Using (\ref{delta 1}) and the properties of $\delta$ functions we obtain

\bea\label{amplitude 1loop 7}
&& \mathcal{M}(\sigma,p,m)  =  \frac{ \lambda^2}{2}    \left[\pi \mathcal{P}\left\{ \int_{ \mathbb{R}^3} \frac{ \, d^3 k}{(2\pi)^4}  \Bigg[  Res[\bar{k}^0_2]+Res[\bar{k}^0_4]-Res[\bar{k}^0_1]-Res[\bar{k}^0_3]     \Bigg]{|}_{\epsilon= 0} \right\} +\right. \nonumber
\\ \nonumber
\\ 
&& - i \mathcal{P}\left\{\int_{ \mathbb{R}^4} \frac{ \, d^4 k}{(2\pi)^4}
\frac{e^{-H\left[\sigma^2\left(k^2-m^2\right)^2\right]}}{k^2 - m^2} \frac{e^{-H\left[\sigma^2\left(\left(k-p\right)^2-m^2\right)^2\right]}}{(k-p)^2 - m^2}\right\}
+ \nonumber\\ 
\\ 
&&
+  \frac{i \left(2\pi\right)^2}{2}    \int_{ \mathbb{R}^4} \frac{ \, d^4 k}{(2\pi)^4}  \delta\left( \left(k-p\right)^2 - m^2\right)\, \delta(k^2 - m^2) \Bigg] \nonumber .
\ee

The first term in (\ref{amplitude 1loop 7}) is real, while the last two terms contribute only to the imaginary part of the amplitude. 
Note that, in the case of the amplitude (\ref{amplitude 1loop 0}) the Cutkosky rules would give, through the prescription (\ref{cutkosky rule}),

\be\label{cutkosky rule one loop}
2 \, i \, \textit{Im}\left\{\mathcal{M}\right\} = -\frac{i \lambda^2}{2}\left(-2\pi i \right)^2   \int_{ \mathbb{R}^4} \frac{ \, d^4 k}{(2\pi)^4}  \delta\left( \left(k-p\right)^2 - m^2\right)\, \delta(k^2 - m^2) \, .
\ee
A simple comparison between (\ref{amplitude 1loop 7}) and (\ref{cutkosky rule one loop}) shows that Cutkosky rules would be  valid if and only if the integral in the second line of (\ref{amplitude 1loop 7}) is identically zero for any $\sigma$, $p$ and $m$. As we will show below, this  is not the case, indeed the Cutkosky rules are no longer valid and the theory is not unitary. 

Let us recast such integral by means of the translation $k\rightarrow k-p/2$, so that  it will be given by the function $I(\sigma,(p^0)^2,m^2)$ defined as



\be\label{definition I 0}
I(\sigma,(p^0)^2,m^2) \equiv \mathcal{P}\left\{\int_{ \mathbb{R}^4} \frac{ \, d^4 k}{(2\pi)^4}
\frac{e^{-H\left[\sigma^2\left(\left(k+\frac{p}{2}\right)^2-m^2\right)^2\right]}}{(k+\frac{p}{2})^2 - m^2 } \frac{e^{-H\left[\sigma^2\left(\left(k-\frac{p}{2}\right)^2-m^2\right)^2\right]}}{(k-\frac{p}{2})^2 - m^2 }\right\}\, ,
\ee
and the condition for the validity of Cutkosky  rules  reads   $I(\sigma,(p^0)^2,m^2)\equiv 0$ for any $\sigma$, $p^0$ and $m^2$.
Note that  the integrand in (\ref{definition I 0}) is even in $p$, so that $I$ must be a function of the scalar $p^2 =(p^0)^2$, as it must in virtue of the Lorentz invariance. Furthermore, 
$I(\sigma,(p^0)^2,m^2)$ is also even in $\sigma$. We stress that the integrand in (\ref{definition I 0}) is also even in $k^0$, but this fact does not determine any specific property of $I(\sigma,(p^0)^2,m^2)$.

Let us derive some general features of this integral. First, note that for  a local theory, corresponding to $\sigma = 0$, this integral is zero. In facts, in this case, the exponential functions in  (\ref{definition I 0}) are  one, and the integrand reduces to the product of the two propagators. By means of elementary partial fraction decomposition, such a product can be recast as

\be\label{definition g}
&&\nonumber\frac{1}{(k+\frac{p}{2})^2 - m^2} \frac{1}{(k-\frac{p}{2})^2 - m^2} = \frac{1}{4 \omega^2} \left[\left(\frac{1}{p^0}-\frac{1}{p^0+2\omega}\right)
\left(\frac{1}{k^0-\frac{p^0}{2}-\omega}-\frac{1}{k^0+\frac{p^0}{2}+\omega}\right) +\right.\\
&&\nonumber\\
&&  \left. \left(\frac{1}{p^0}-\frac{1}{p^0-2\omega}\right)
\left(\frac{1}{k^0-\frac{p^0}{2}+\omega}-\frac{1}{k^0+\frac{p^0}{2}-\omega}\right)
\right] ,
\ee
where $\omega=\omega(\vec{k},\epsilon=0)\equiv\omega(\vec{k})$, indeed, the integration in the $k^0$ variable in (\ref{definition I 0}) reduces to a sum of integrals of the type

\be\label{integral local case 1}
\mathcal{P}\left\{\int_{ \mathbb{R}}  d k^0
 \frac{1}{k^0 \pm \left(\frac{p^0}{2} \pm \omega(\vec{k})\right)} \right\} = 0 \, ,
\ee
that are all null, because the principal part is taken integrating in an interval symmetric with respect to the pole. Indeed, in the local case, $I(\sigma=0,p^0,m^2)=0$ by symmetry. This implies that the second term in (\ref{amplitude 1loop 7}) is zero for $\sigma=0$, and the imaginary part of the amplitude coincides with the last term, so that Cutkosky rules are valid and the local theory is unitary. 

In conclusion, the local limit implies that   $I(\sigma, p^0,m^2)$ must be such that
\be\label{limit 0+}
\lim\limits_{\sigma \rightarrow 0} I(\sigma, p^0,m^2) = 0 \,,  \qquad \sigma \in \mathbb{R} , \forall \, p^0, m^2 \, .
\ee
Moreover, by means of the change of variables $k\rightarrow k/\sqrt{|\sigma|}$ in the defining expression (\ref{definition I 0}) it is easy to show that 

\be\label{Limit I 2}
I(\sigma, (p^0)^2,m^2)= I(\sigma = 1,|\sigma|(p^0)^2,|\sigma| m^2) = \phi(|\sigma| \left(p^0\right)^2,|\sigma| m^2)), 
\ee
so that $I$ depends on $\sigma$  through the variables $|\sigma| \left(p^0\right)^2$ and $|\sigma| m^2$.\footnote{This is a consequence of the fact that $I$ is dimensionless.} From (\ref{Limit I 2}) we already see that $I$ is not analytic in $\sigma =0$, since it depends on $|\sigma|$, which is not an analytic function of $\sigma$. 
Form (\ref{definition I 0}) it is also easy to see that 
\be\label{Limit I 1}
I(\sigma, (a p^0)^2,m^2)= I(a^2 \sigma,(p^0)^2,m^2/a^2)\, , 
\ee
so that, using (\ref{limit 0+}) one has
\be
\lim\limits_{a \rightarrow 0} I(\sigma, (a p^0)^2,m^2) = \lim\limits_{a \rightarrow 0} I(a^2\sigma, (p^0)^2,m^2/a^2)= 0 \, \qquad \forall \quad  m \, ,
\ee
which means that
\be
I(\sigma, (p^0)^2=0,m^2) = \phi(|\sigma| (p^0)^2= 0, |\sigma| m^2) = 0 \, \qquad \forall \quad  m \, .
\ee
These relations implies that $I$ must have the form

\be
I(\sigma, (p^0)^2,m^2) =  f(|\sigma| (p^0)^2) \times \psi(|\sigma| (p^0)^2, |\sigma| m^2), \, \text{with }  f(0)=0, \psi(0,0)\neq 0 .
\ee
In the Appendix \ref{appendix} we will give an example of an integral function with properties similar to those of $I(\sigma, (p^0)^2,m^2)$ that can be calculated explicitly.

Regardless the details of the functions $f(|\sigma| (p^0)^2) $ and $\psi(|\sigma| (p^0)^2),|\sigma|m^2)$, we can prove that $I(\sigma,(p^0)^2,m^2)\not\equiv 0$. Let us express $I$ as a function of $\alpha = \sigma^2 \in \mathbb{R}^+_0$, so that

\be\label{definition I 3}
&&\tilde{I}(\alpha,(p^0)^2,m^2) \equiv I(\sqrt{\alpha},(p^0)^2,m^2)= \phi\left(\sqrt{\alpha}(p^0)^2,\sqrt{\alpha}m^2\right)=\\ \nonumber
\\\nonumber
&&= \mathcal{P}\left\{\int_{ \mathbb{R}^4} \frac{ \, d^4 k}{(2\pi)^4}
\frac{e^{-H\left[\alpha \left(\left(k+p/2\right)^2-m^2\right)^2\right]-H\left[\alpha\left(\left(k-p/2\right)^2-m^2\right)^2\right]}}{\left((k-p/2)^2 - m^2 + i \epsilon\right) \left((k+p/2)^2 - m^2 + i \epsilon\right)}\right\} \, .
\ee
Since $H[0]=0$ (see the discussion below (\ref{phin2})), it must be $H[z]= z \left(\sum_{n=0}^{\infty}c_n z^n\right)$, which gives $H^\prime[0]= c_0 \equiv H^\prime_0$. In order to show that $\tilde{I}(\alpha,(p^0)^2,m^2)\not\equiv 0$ it is sufficient to show that its first derivative is not identically zero. One has

\be\label{definition I 4}
&&\partial_\alpha \tilde{I}(\alpha,(p^0)^2,m^2)= - \mathcal{P}\left\{\int_{ \mathbb{R}^4} \frac{ \, d^4 k}{(2\pi)^4}
\frac{e^{-H\left[\alpha\left(\left(k+p/2\right)^2-m^2\right)^2\right]-H\left[\alpha\left(\left(k-p/2\right)^2-m^2\right)^2\right]}}{\left((k-p/2)^2 - m^2 + i \epsilon\right) \left((k+p/2)^2 - m^2 + i \epsilon\right)}\right. \times\\ \nonumber
\\\nonumber
&& \left.\left[H^\prime\left[\alpha\left(\left(k+\frac{p}{2}\right)^2-m^2\right)^2\right]\left(\left(k+\frac{p}{2}\right)^2-m^2\right)^2+H^\prime\left[\alpha\left(\left(k-\frac{p}{2}\right)^2-m^2\right)^2\right]\left(\left(k-\frac{p}{2}\right)^2-m^2\right)^2\right]  \right\},
\ee
so that

\be\label{definition I 10}
&&\lim\limits_{\alpha\rightarrow0^+}\vert \partial_\alpha \tilde{I}(\alpha,(p^0)^2,m^2) \vert= \\
\nonumber
\\
&&\nonumber \Bigg| H^\prime_0 \mathcal{P}\left\{\int_{ \mathbb{R}^4} \frac{ \, d^4 k}{(2\pi)^4}
\frac{\left(\left(k+p/2\right)^2-m^2\right)^2+\left(\left(k-p/2\right)^2-m^2\right)^2}{\left((k-p/2)^2 - m^2 + i \epsilon\right) \left((k+p/2)^2 - m^2 + i \epsilon\right)}\right\} \Bigg| = + \infty,
\ee
as the last integral is infinite, since its integrand goes to $1$ when $k^2 \rightarrow \pm \infty$. 

Indeed, (\ref{definition I 10}) implies that $I(\sigma,(p^0)^2,m^2)=\tilde{I}(\alpha,(p^0)^2,m^2)\not\equiv 0$, therefore the second term in (\ref{amplitude 1loop 7}) is not identically zero, and the imaginary part of the complex amplitude (\ref{amplitude 1loop 0}) is not given by the Cutkosky rules as in (\ref{cutkosky rule one loop}). This concludes our proof of the unitarity breaking  of the Minkowskian non-local field theory.

We just mention that this proof generalizes straightforwardly to the case $H[z]= z^m \left(\sum_{n=0}^{\infty}c_n z^n\right)$, as one  can define $\alpha= \sigma^{2m}$ and proceed in the same way to show that $|\partial_\alpha I((\alpha)^{1/2m},(p^0)^2,m^2)|\rightarrow + \infty$ for $\alpha\rightarrow0^+$. Indeed, unless $H\equiv 0$, that corresponds to the local theory, $I(\sigma,(p^0)^2,m^2)$ cannot be identically zero.

\section{One-loop diagram in the Euclidean non-local theory}\label{Section one loop Euclid}

In order to clarify the differences between the Minkowskian and Euclidean theories, in this section we consider the same diagram in Fig.\ref{Fig2} in the Euclidean case, and show that the Cutkosky rules give the right result (\ref{cutkosky rule one loop}) for the imaginary part of the complex amplitude. However, we remand the reader to the references \cite{unitarity1,unitarity2,unitarity3} for more details.

The amputated amplitude for the one-loop diagram in Fig.\ref{Fig2} in the Euclidean theory is 

\begin{equation}\label{amplitude euclid 1loop 0}
\mathcal{M}(\sigma,p,m,\epsilon) = -  \frac{i \lambda^2}{2} \, \int_{
\mathcal{I} \times	\mathbb{R}^3} \frac{ \, d^4 k}{(2\pi)^4}
\frac{e^{-H\left[\sigma\left(k^2-m^2\right)\right]}}{k^2 - m^2 + i \epsilon} \frac{e^{-H\left[\sigma\left(\left(k-p\right)^2-m^2\right)\right]}}{(k-p)^2 - m^2 + i \epsilon} ,
\end{equation}
where   $p = p_1+p_2=p_3+p_4$, and $p_1, p_2, p_3,p_4$ are the external momenta. The external energy $p^0$ in (\ref{amplitude euclid 1loop 0}) is assumed to be purely imaginary, and the $k^0$ integration is performed along the imaginary axis $\mathcal{I}$ of the complex $k^0$ plane for $k^0$ going from $- i \infty$ to $+ i \infty$. The physical amplitude is then obtained by analytic continuation of (\ref{amplitude euclid 1loop 0}) to real energies $p^0 \in \mathbb{R}^+_0$.

When the theory is defined in Euclidean signature, unitarity is ensured assuming that $\exp\left(H[z]\right)$ is an entire function  (analytic with no poles except for $|z|=\infty$) without zeros at finite $z$ \cite{unitarity1,unitarity2,unitarity3}. Moreover, to enforce the convergence of scattering amplitudes and achieve the renormalizability of NLQG, it is assumed that $z^\gamma \exp\left(-H\left[z\right]\right)\rightarrow 0$ when $z\rightarrow - \infty$ for some $\gamma >0$.
Again, we assume that $H\left[0\right]=0$ and we  set $\vec{p}=0$ without loss of generality.

Before explaining how this analytic continuation is obtained, let us clarify some detail. In the Euclidean case, the convergence of the integral requires that $H[z]$  must be a function of $\sigma\left(k^2-m^2\right)$ such that $H[z]\rightarrow +\infty$ for $z\rightarrow -\infty$, so that the exponential form factor goes to zero for $k \rightarrow \infty$ in any direction of $\mathbb{I}\times \mathbb{R}^3$. For instance, in the Euclidean theory both the functions $H= - \sigma\left(k^2-m^2\right)$ and $H=\left(\sigma\left(k^2-m^2\right)\right)^2$ are admissible, while the first would give a divergent amplitude in the Minkowskian theory. However, even though the second form factor is admitted in both the Euclidean and Minkowskian theories, the latter is not unitary, as we have shown in Section \ref{Section one loop Mink}. It is worth to stress that the integrals in the Euclidean and Minkowskian theories are not connected by a Wick rotation, since the exponential form factors in (\ref{amplitude 1loop 0}) and (\ref{amplitude euclid 1loop 0}) have an essential singularity at infinity in the complex $k^0$ plane, indeed the integration on the infinite arcs,  that would connect the integration paths  in the two theories (the imaginary and real  axis in the Euclidean and Minkowskian cases respectively), are not zero.

Again, a key feature to ensure the unitarity of the Euclidean theory is that the form factor $\exp\left(H\left[z\right]\right)$ has no poles in the finite complex $z$ plane. Indeed, the poles of the integrand in (\ref{amplitude euclid 1loop 0}) are only those of the two propagators given in (\ref{poles 1}-\ref{poles 2}), which are the same of both the local and the Minkowskian non-local theories, so the singularity structure of the amplitudes is the same in all such theories.

\begin{figure}
	\begin{center}
		\hspace{-1cm}
		\includegraphics[height=5cm]{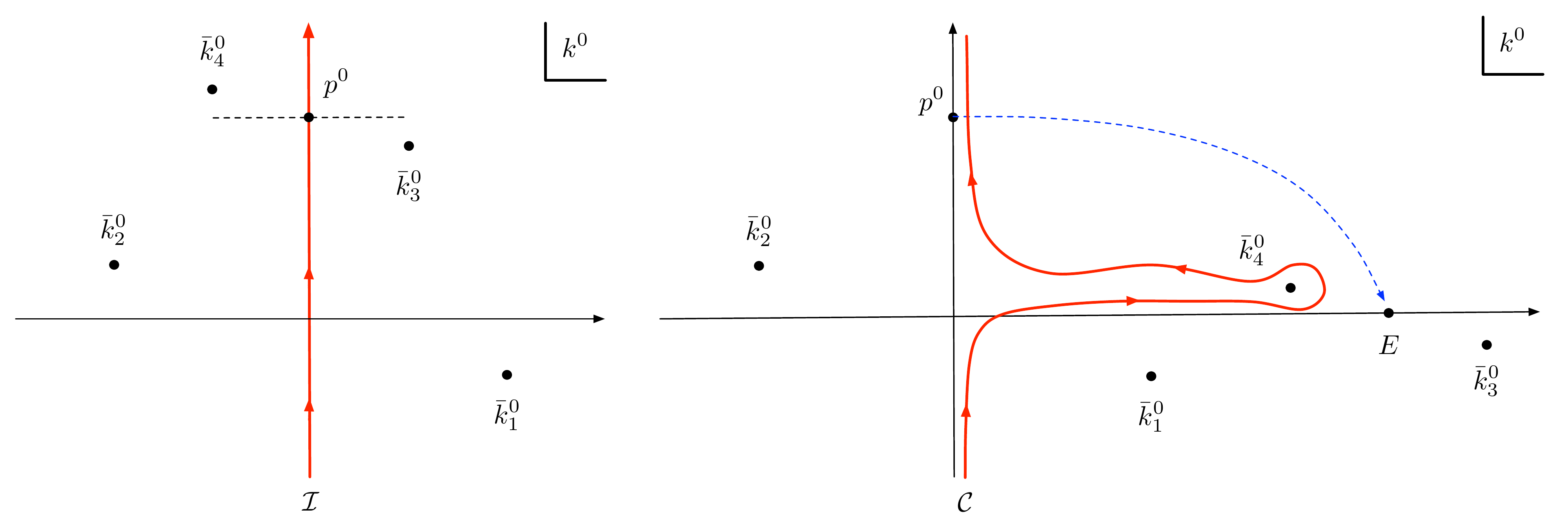}
	\end{center}
\caption{(Left) We plot the poles $\bar{k}^0_1, \bar{k}^0_2,\bar{k}^0_3,\bar{k}^0_4$ on the complex $k^0$ plane,  when $p^0$ is purely imaginary. (Right) We plot the same poles for $p^0$ real and positive. Since $\bar{k}^0_1$ and $\bar{k}^0_2$ do not depend on $p^0$, their positions do not change when $p^0$ becomes real. On the contrary, $\bar{k}^0_3$ and $\bar{k}^0_4$ move to the right, and $\bar{k}^0_4$ passes through the imaginary axis $\mathcal{I}$ for some values of $\vec{k}$. We also plot the contour $\mathcal{C}$, which is obtained deforming $\mathcal{I}$ around $\bar{k}^0_4$.}	\label{Fig3old}
\end{figure}

Let us now describe how to obtain the analytic continuation of (\ref{amplitude euclid 1loop 0}) to real external energies. When $p^0$ is purely
imaginary, the two poles $\bar{k}^0_{2}$ and $\bar{k}^0_{4}$ are  at
the left of $\mathcal{I}$, while $\bar{k}^0_{1}$ and $\bar{k}^0_{3}$ are  at the right of $\mathcal{I}$, see Fig.\ref{Fig3old} (left). When $p^0$ is moved to real and positive values, the two poles $\bar{k}^0_{3}$ and $\bar{k}^0_{4}$ move to the right, and 
$\bar{k}^0_{4}$ passes through the imaginary axis for  values of the loop momenta $\vec{k}$ such that $|\vec{k}|^2< \Re\{p^0\}^2-m^2$ for $\Re\left\{p^0\right\}> m$, see Fig.\ref{Fig3old} (right).
Indeed, the analytic continuation of (\ref{amplitude euclid 1loop 0}) to real $p^0$ is obtained deforming the integration contourn $\mathcal{I}$ around the pole $\bar{k}^0_{4}$, obtaining the contour $\mathcal{C}$, so that

\begin{equation}\label{amplitude euclid 1loop 1}
\mathcal{M}(\sigma,p,m,\epsilon) = -  \frac{i \lambda^2}{2} \, \int_{
	\mathcal{C} \times	\mathbb{R}^3} \frac{ \, d^4 k}{(2\pi)^4}
\frac{e^{-H\left[\sigma\left(k^2-m^2\right)\right]}}{k^2 - m^2 + i \epsilon} \frac{e^{-H\left[\sigma\left(\left(k-p\right)^2-m^2\right)\right]}}{(k-p)^2 - m^2 + i \epsilon} ,
\end{equation}
for $p^0 \in \mathbb{R}^+_0$. Finally, using the residue theorem and sending $\epsilon\rightarrow 0$ one has

\bea\label{amplitude euclid 1loop 3}
&& \mathcal{M}(\sigma,p,m) \equiv \lim_{\epsilon \rightarrow 0} \mathcal{M}(\sigma,p,m,\epsilon)  =  \frac{ \lambda^2}{2}    \left[- i  \int_{\mathcal{I}\times \mathbb{R}^3} \frac{ \, d^4 k}{(2\pi)^4}
\frac{e^{-H\left[\sigma\left(k^2-m^2\right)\right]}}{k^2 - m^2} \frac{e^{-H\left[\sigma\left(\left(k-p\right)-m^2\right)^2\right]}}{(k-p)^2 - m^2 }
\right.\nonumber 
\\ \nonumber
\\ 
&&+ 2 \pi \lim_{\epsilon \rightarrow 0} \int_{ \mathbb{R}^3} \frac{ \, d^3 k}{(2\pi)^4}   Res[\bar{k}^0_4] \Bigg] ,
\ee
where $Res[\bar{k}^0_4] $ is given by the last of (\ref{residues}) with the replacement 
\mbox{ $H[\sigma^2\left(k^2-m^2\right)^2]\rightarrow H[\sigma\left(k^2-m^2\right)]$}.
Using (\ref{delta 1}) and the properties of the $\delta$ function, one finally has

\bea\label{amplitude euclid 1loop 4}
&& \mathcal{M}(\sigma,p,m)  = \frac{ \lambda^2}{2}    \left[ \mathcal{P}\left\{ \int_{ \mathbb{R}^3} \frac{ \, d^3 k}{(2\pi)^3}   Res[\bar{k}^0_4] {|}_{\epsilon= 0}\right\}  +
\right. 
\\ \nonumber
\\ \nonumber
&&  - i  \int_{\mathcal{I}\times \mathbb{R}^3} \frac{ \, d^4 k}{(2\pi)^4}
\frac{e^{-H\left[\sigma\left(k^2-m^2\right)\right]}}{k^2 - m^2} \frac{e^{-H\left[\sigma\left(\left(k-p\right)-m^2\right)^2\right]}}{(k-p)^2 - m^2 }+  \\ \nonumber
\\ 
&& \frac{i\left(2\pi\right)^2}{2}    \int_{ \mathbb{R}^4} \frac{ \, d^4 k}{(2\pi)^4}  \delta\left( \left(k-p\right)^2 - m^2\right)\, \delta(k^2 - m^2)
\Bigg].\nonumber
\ee
Since the first two terms in (\ref{amplitude euclid 1loop 4}) are real, only the last term contributes to the imaginary part of the amplitude, so that the Cutcosky rule will give the correct result (\ref{cutkosky rule one loop}) for $\textit{Im}\left\{\mathcal{M}\right\}$, according to the fact that the Euclidean theory is unitary \cite{unitarity1,unitarity2,unitarity3}.

Comparing (\ref{amplitude euclid 1loop 4}) with (\ref{amplitude 1loop 7}) we note that the key difference between the  Minkowskian and the Euclidean cases is that the second term in (\ref{amplitude 1loop 7}) and (\ref{amplitude euclid 1loop 4}) is respectively purely imaginary and real, so in the first case it gives an extra contribution to the imaginary part of $\mathcal{M}$ that brakes the Cutkosky rules and spoils the unitarity of the theory.

\section{Conclusions}\label{Section conclusions}

Considering the case of the simple one-loop diagram in Fig. \ref{Fig2}, in this paper we have shown that, when the field theory (\ref{phin2}) is defined in  Minkowskian signature, the imaginary part of the complex amplitudes is no longer given by the Cutkosky cutting rules, but it contain extra terms as in the second line of (\ref{amplitude 1loop 7}). Since Cutkosky rules are at the basis of the unitarity of a quantum field theory, we conclude that the Minkowskian field theories are not unitary.

\appendix

\section{}\label{appendix}

As an example of what we stated in Section \ref{Section one loop Mink} about $I(\sigma,(p^0)^2,m^2)$, we consider the following one-dimensional model integral, that has the same symmetries of (\ref{definition I 0}):

\be\label{example A}
A(\sigma,p)= \mathcal{P} \left\{ \int_\mathbb{R} dx\, \frac{1}{1+\sigma^2 (x-p)^2}\frac{1}{1+\sigma^2 (x+p)^2} \left(\frac{1}{x-p}-\frac{1}{x+p}\right)\, S(p)\right\}  \, ,
\ee
for $\sigma \in \mathbb{R}$, where $S(p)$ is the sign of $p$, so that one has the symmetries $A(\sigma,p) = A(\sigma,-p)=A(-\sigma,p)$. Such symmetries, together with the fact that $A(\sigma=0,p) = 0$ and that $A(\sigma,p)$ is dimensionless, implies that it must be $A(\sigma,p) \propto |\sigma p|^\gamma$ with $\gamma>0$. An explicit calculation gives

\be\label{example A 2}
A(\sigma,p)= - 3  \pi |\sigma p|  \left[1-5 (|\sigma p|)^2+21 (|\sigma p|)^4)+O(|\sigma p|)^6\right] \neq 0 \, ,
\ee
In analogy to (\ref{definition I 3}), if we define $\alpha = \sigma^2 \in \mathbb{R}^+_0$, so that
\be\label{example A 3}
A(\sqrt{\alpha},p)= - 3  \pi \sqrt{\alpha} |p|  \left[1-5 \alpha (p)^2+21 \alpha^2(p)^4)+O(\sqrt{\alpha} p)^6\right] \neq 0 \, 
\ee
we get

\be\label{example A 4}
\partial_\alpha A(\sqrt{\alpha},p) = - \frac{3  \pi}{2} \frac{|p|}{\sqrt{\alpha}}+ O(|p|) \rightarrow - \infty \, ,
\text{for}  \,\, \alpha \rightarrow 0^+ ,  
\ee
that is the analogue of (\ref{definition I 10}).


\end{document}